\documentclass[twocolumn,prd,showpacs,superscriptaddress,groupedaddress,preprintnumbers,nofootinbib]{revtex4-1}
\pdfoutput=1
\usepackage{graphicx,amsfonts,amsmath,amssymb,epsfig,color, mathrsfs,latexsym,url,wasysym,float,xfrac}
\newcommand{\be}{\begin{equation}}
\newcommand{\ee}{\end{equation}}
\newcommand{\nn}{\nonumber}
\newcommand{\ba}{\begin{eqnarray}}
\newcommand{\ea}{\end{eqnarray}}
\newcommand{\mpl}{m_{\rm Pl}}

\newcommand{\Hi}{H_{\rm inf}}

\newcommand{\Hrad}{H_{\rm rad}}
\newcommand{\Trad}{T_{\rm rad}}

\newcommand{\ls}{\lesssim}
\newcommand{\gs}{\gtrsim}

\begin{document}

\title{Breathing Comoving Hubble: \\ Initial Condition and Eternity in view of the trans-Planckian Censorship Conjecture} 

\author{Mahdi Torabian$^*$}
\affiliation{Department of Physics, Sharif University of Technology, Azadi Ave, 11155-9161, Tehran, Iran} 

\begin{abstract}
In this paper we put forward the idea that the comoving Hubble horizon undergoes multiple stages of contraction ({\it a.k.a.} inflationary phase) and expansion. The observable inflation, that produces  the CMB anisotropies and generates primordial gravitational waves,  follows and is followed by multiple early and late inflations. The trans-Planckian censorship conjecture restricts the duration of each inflationary phases and determines their Hubble rates. Early inflations could start immediately after the universe emerges from the Planck era. It alleviates the initial condition problem for the lower-scale observable inflation. Late inflations collectively assist the observable inflation to accommodate the present horizon. Moreover, it makes eternal inflation possible and consistent with obserbvations. 
\end{abstract}

\preprint{SUT/Physics-nnn}
\maketitle

\subsection*{I. Introduction} 
The paradigm of cosmic inflation was proposed to alleviate the fine-tuning problem in the initial conditions of the standard big bang model \cite{Starobinsky:1980te,Guth:1980zm,Linde:1981mu,Albrecht:1982wi}.   During inflation, the universe experiences an accelerated expansion so that its comoving Hubble horizon shrinks and physical scales are stretched. Without inflation, the present horizon would be composed of around $10^6$ causally disconnected patches at the time of recombination and about $10^{27}$ parts at the time of nucleosynthesis with no correlated past histories. It is in direct tension with the observed large scale homogeneities from the CMB and the LSS measurements. In principle, the big bang model can be applied immediately after the universe has emerged from the Planck epoch. Then the current horizon consists of around $10^{87}$ isolated pieces at the Planck time. The natural expectation is that inflation instantly takes over, provides the initial kick for Hubble expansion and expands a Planck size smooth patch to make the observed universe.  

Furthermore when quantum effects are considered, inflations makes genuine predictions. It amplifies vacuum quantum fluctuations in scalar and gravitational fields \cite{Mukhanov:1981xt,Mukhanov:1982nu,Starobinsky:1982ee}. Physical modes are stretched to superhorizon scales, classicalized and freeze. They re-enter the horizon at later stages of the cosmic evolution. Therefor, inflation predicts small inhomogeneities and primordial gravitational waves on the background.  The former is actually observed as temperature anisotropies on the CMB. In the last 30 years, these anisotropies has been measured with high accuracy and provide a solid support for inflationary dynamics \cite{Bennett:1996ce,Hinshaw:2012aka,Akrami:2018vks}. The latest results from the {\sl Planck} satellite, in particular, favors a lower scale inflation with Hubble rate $\Hi\sim10^{14}$ GeV \cite{Akrami:2018odb}
. The delayed phase of inflation, besides losing the beginning bang, reintroduce the initial condition problem \cite{Goldwirth:1989pr,Goldwirth:1991rj,Ijjas:2013vea}
. It needs to overcome inhomogeneities propagating inward the horizon and preclude inflation. The initial smooth patch of size around $\Hi^{-1}$ must span over $10^4$ to $10^6$ parts at the Planck time, for matter and radiation intervening epochs respectively (see \cite{Guth:2013sya,Linde:2014nna,Chowdhury:2019otk}  for supporting arguments). 

Moreover under general assumptions, quantum fluctuations can take over classical dynamics and prevents inflation from termination \cite{Linde:1986fd,Starobinsky:1986fx,Aryal:1987vn,Vilenkin:1999pi,Guth:2000ka,Linde:2015edk}. Once inflation began it never ends completely all over. Eternal inflation can populate vacua in the string landscape by the so-called pocket universes. Although there is no direct experimental handle to this prediction, this picture of multiverse offers a solution to fine-tuning of low energy parameters. 

Inflation is a quasi-de Sitter (dS) state of the universe and is best understood in terms of an effective field theory (EFT) below the Planck scale. In general, it is difficult to incorporate the dS space in a theory of quantum gravity; both technically and conceptually. For instance, there are far fewer constructions of dS solutions than anti dS/Minkowski ones within string theory. It motivates to prepare a set of criteria that every low energy EFT (that induces positive cosmological constant) must fulfill in order to have UV completion in a quantum gravity \cite{Vafa:2005ui,ArkaniHamed:2006dz,Ooguri:2016pdq}. Examples of these conditions, called the swampland conjectures, are the dS conjecture \cite{Obied:2018sgi} (and a milder version \cite{Garg:2018reu,Ooguri:2018wrx}
) and the distant conjecture \cite{Ooguri:2006in,Klaewer:2016kiy}
. There are no proofs yet but they have strong support form many examples in string theory (we refer the reader to a review \cite{Palti:2019pca} of the vast literature). 

The swampland conjectures have important consequences for inflation as the shape of potential and the field excursion is highly constrained (the literature is too extensive to be covered here, instead we refer to the pioneering studies \cite{,Agrawal:2018own,Achucarro:2018vey,Kehagias:2018uem,Kinney:2018nny,Brahma:2018hrd,Das:2018hqy}). Moreover, there are arguments that eternal inflation is not allowed \cite{Rudelius:2019cfh,Dvali:2018fqu,Dvali:2018jhn} or is marginally happens but not consistent with observation \cite{Matsui:2018bsy,Dimopoulos:2018upl,Kinney:2018kew,Andrzejewski:2018pwq,Wang:2019eym} as it cannot be followed by a normal inflation.

Recently, another swampland condition, the trans-Planckian censorship conjecture (TCC),  was proposed in \cite{Bedroya:2019snp}. Quantum modes with wavelength smaller than the Planck length belong to the theory of quantum gravity and cannot be describe by a low energy EFT below Planck scale. In an expanding background these modes are stretched and if they extend beyond the horizon then they freeze and classicalize. If these modes subsequently re-enter the horizon, they affect classical spacetime which is described by a low energy EFT like general relativity. The TCC proposes that an EFT never classicalizes sub-Planckian modes otherwise belong to the swampland. Immediate implications follow for the present status of the universe and during primordial inflation \cite{Bedroya:2019tba}. In particular it constrain the number of e-folds or equivalently the Hubble rate (at the end) of expansion as
\be e^{N}\equiv\frac{a_{{\rm end}}}{a_{{\rm ini}}}<\frac{\mpl}{H_{{\rm end,inf}}},\ee
where $a$ it the scale factor and $\mpl\simeq 2.4\times 10^{18}$ GeV is the reduced Planck mass.   
On the other hand, a minimum number e-folds are needed to explain the present horizon. If we assume a standard scenario with a single inflation, followed by an epoch with EoS parameter $w$ and Hubble parameter $H_1$ at the end of that epoch, then we find
\ba\label{master} 
 \Hi^{2-\sfrac{2}{3(1+w)}}\!\!<\! \mpl H_0 T_0^{-1}\Trad H_1^{-\sfrac{2}{3(1+w)}}.
\ea
The temperature at the onset of radiation domination $\Trad$ is greater than about 10 MeV to initiate nucleosynthesis. For radiation and matter domination following inflation we find
$\Hi<0.1\ {\rm GeV}$ and $\Hi< 50\ {\rm GeV}\cdot(10{\rm MeV}/\Trad)^{\sfrac{1}{4}}$,  respectively (see \cite{Cai:2019hge,Tenkanen:2019wsd,Das:2019hto,Mizuno:2019bxy,Brahma:2019unn,Draper:2019utz,Dhuria:2019oyf,Torabian:2019zms,Cai:2019igo,Schmitz:2019uti,Kadota:2019dol,Berera:2019zdd,Brahma:2019vpl,Goswami:2019ehb,Okada:2019yne,Lin:2019pmj,Li:2019ipk,Kehagias:2019iem,Saito:2019tkc} for further elaborations and refinements). Low scale Hubble parameter implies huge amount of fine-tuning in initial condition of inflation. The smooth patch at the onset of inflation includes $10^{17}$ or $10^{30}$ similar parts at the Planck time, for matter and radiation epochs respectively. Moreover, it implies no primordial gravitational waves will be detected in future observations. 

Strong pieces of evidence imply that the universe has experienced at two stages of inflation. It is  possible that the universe has followed many stages of inflation. In \cite{Torabian:2019zms} we put forward the idea that the upper bound \eqref{master}  can be relaxed multiple inflation happens 
\ba\label{master-2} \Hi^{2-\sfrac{2}{3(1+w)}}<&&\ \mpl H_0 T_0^{-1}\Trad  H_1^{-\sfrac{2}{3(1+w)}} \cr &&\qquad\qquad \times e^{\sum_iN_i}{\textstyle\prod_i}\Big(\frac{H_i}{H_{i+1}}\Big)^{\sfrac{2}{3(1+w_i)}},\ \ \ea
where $w_i$ is the EoS parameters of intervening epoch $i$ with final expansion rate $H_i$. The observable inflation has imprints on the CMB and the LSS observables and the following ones assist to explain the present horizon (multiple inflation was previously studied by different motivations \cite{Adams:1997de,Tetradis:1997kp,Burgess:2005sb,Kawasaki:2010ux,Cicoli:2014bja,Dimopoulos:2016yep}). Moreover, it is possible that there are epochs of inflations preceding the observable one that prepare the smooth patch and alleviate the initial condition for it to happen. Additionally, eternal inflation is permitted as the follow-up slow-role inflations are made consistent with observations. 

The initial condition and the eternity are the most debated subjects in the inflationary paradigm. In this note, we study these issues in the light of the TCC in the framework of multiple inflation. A generic scenario is illusterated in Fig. 1 and will be discussed in the following sections.
The paper is structured as follows. In section II we see that a high scale observable inflation is consistent with the TCC. In section III we solve the initial condition problem for the observable inflation through introducing pre-inflations. In section IV we study power-law inflations precede and follow the plateau-like observable inflation. Finally, we conclude in section V. 
\vspace*{0mm}\begin{center}\begin{figure}[t!]\includegraphics[scale=.49]{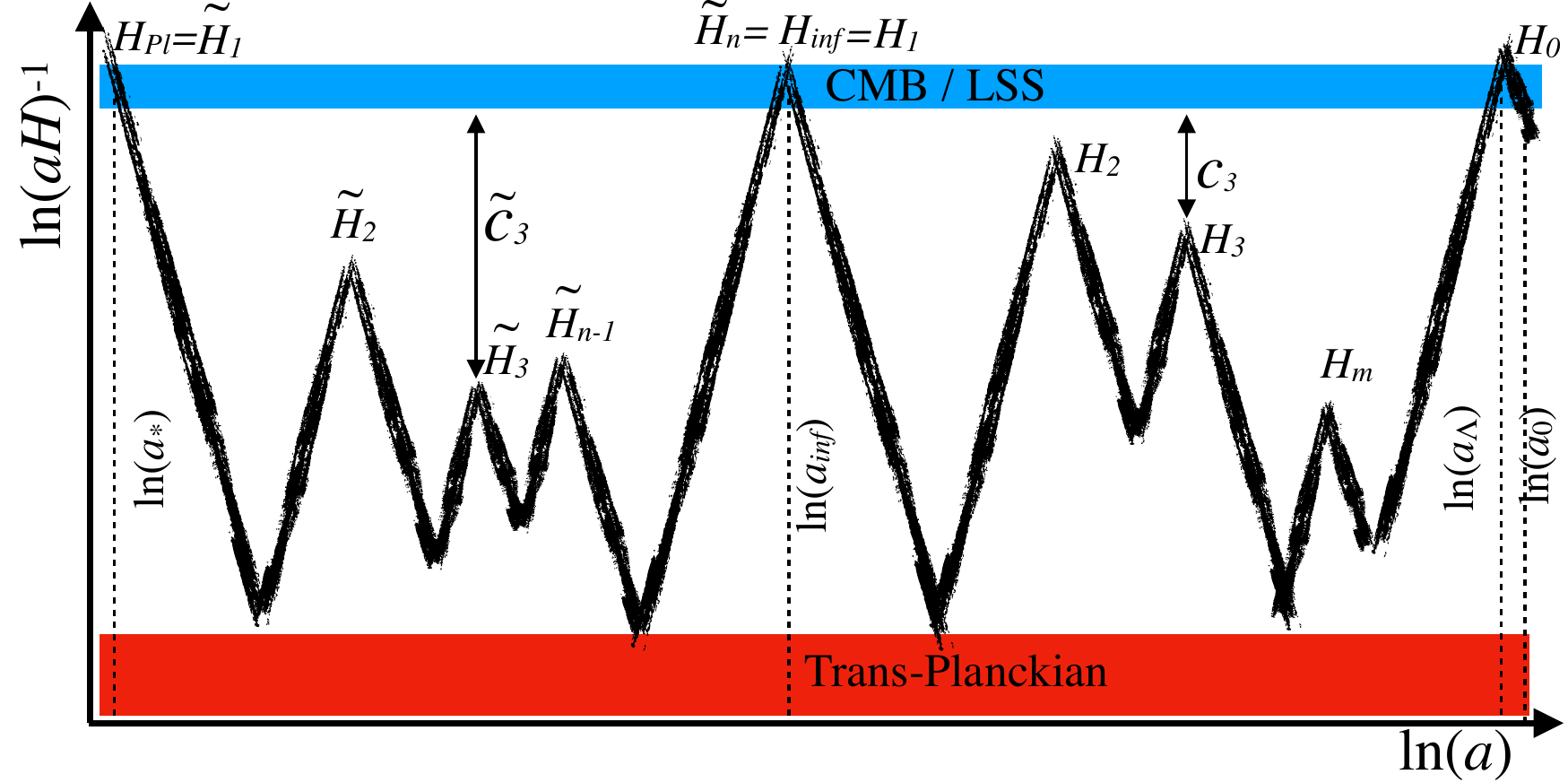}\caption{A generic scenario with breathing comoving Hubble horizon. The blue band is the observational window and the red floor is provided by the TCC.}\end{figure}\end{center} \vspace*{-16mm}

\subsection*{II. Observable inflation at high-scale}
In this section we study the case of high-scale observable inflation. First we compute the total number of e-folds needed to explain the horizon. Then, applying the TCC, we compute the Hubble rate for each inflationary epoch. In the end, we present a simple scenario.

\subsubsection*{The present horizon with multiple inflation}
To address the the horizon problem, the present comoving Hubble horizon be inside the comoving horizon the the onset of observable inflation. If numerous inflations with scale $H_i$ happen then we find 
\ba 1&&\ge\!\frac{a_{{\rm ini},1}\Hi}{a_0H_0}\!=\!\frac{a_{{\rm ini},1}}{a_{{\rm end},1}}\frac{a_{{\rm end},1}}{a_{{\rm ini},2}}\cdots\frac{a_{{\rm ini},m-1}}{a_{\rm end,m}}\frac{a_{{\rm end},m}}{a_{\rm rad}}\frac{a_{\rm rad}}{a_0}\frac{\Hi}{H_0}\cr
\!&&=\!e^{-\sum_{i=1}^mN_i}\frac{T_0}{\Trad}\frac{\Hi}{H_0}\frac{a_{{\rm end},m}}{a_{\rm rad}}\!\\
&&=\! e^{-\sum_{i=1}^mN_i}\!\frac{T_0}{\Trad}\!\frac{\Hi}{H_0}\Big(\frac{\Hrad}{H_m}\Big)^{\frac{2}{3(1+w_m)}}\prod_{i=1}^{m-1}\!\!\Big(\frac{H_{i+1}}{H_i}\Big)^{\frac{2}{3(1+w_i)}},\nn\ea
where index 1 indicates the observable inflation ($\Hi=H_1$) and $m$ is the last one.
In the third line we assume that a single stage with EoS parameter $w_i$ intervene between two inflations. For simplicity, we assume $w_i=w$ then the total number of e-folds $N_{\rm tot}=\sum_{i=1}^mN_i$ satisfies
\ba\label{min-e-folds} N_{\rm tot}\ge\ln\Big(\frac{T_0}{H_0}\Big)+\frac{1}{3(1+w)}\ln\big(\Trad^{1-3w}\Hi^{1+3\omega}\mpl^{-2}\big).\quad \ea


\subsubsection*{The upper bound on the Hubble parameters}
We parametrize the ratio of the comoving horizon at the onset if each inflation to the present horizon as
\ba c_i&=&\frac{a_{{\rm ini},i}H_i}{a_0H_0}.\ea
The first inflation is assumed to be the observable one. It generates the CMB anisotropies and the gravitational waves. Thus, $c_1\le1$. On the other hand, the succedding inflations must not jeopardize the success of CMB observations which are in the range
 \be 10^{-3}\ls\frac{k}{a_0H_0}\ls 1.\ee
Therefore,  $c_i\ge10^3$ for $i=2\dots,m$. Moreover, if we happen to have a homogeneous patch before the onset of the first inflation, as the comoving Hubble volume never surpasses the initial one, the following inflations will follow without any further tuning. 

For the final ($m$th) stage of inflation we find that
\ba\label{m-th} c_m&=&\frac{a_{{\rm ini},m}H_m}{a_0H_0}=\frac{a_{{\rm ini},m}}{a_{{\rm end},m}}\frac{a_{{\rm end},m}}{a_{\rm rad}}\frac{a_{\rm rad}}{a_0}\frac{H_{m}}{H_0}\cr
&=& e^{-N_m}\Big(\frac{\Hrad}{H_m}\Big)^{\frac{2}{3(1+w_m)}}\frac{T_0}{\Trad}\frac{H_{m}}{H_0},\qquad\ea
where in second line we assumed a single stage epoch. Applying the TCC inequality, we find an upper bound on the Hubble rate $H_m$ as
\ba\label{Hm} \Big(\frac{H_m}{\mpl}\Big)^{2+\frac{2}{1+3w_{m-1}}}\!<\!\big(c_mH_0T_0^{-1}\big)^{\frac{3(1+w_{m})}{1+3w_{m}}}
\Big(\frac{\mpl}{\Trad}\Big)^{\frac{1-3w_m}{1+3w_m)}}.\ \quad\ea
For the preceding inflations for $i=m\!-\!1,m\!-\!2,\dots,2,1$, we find
\ba c_{i}&=&\frac{a_{{\rm ini},i}H_{i}}{a_0H_0}= c_{i+1}\frac{a_{{\rm ini},i}H_{i}}{a_{{\rm ini},i+1}H_{i+1}}\\
&=&c_{i+1}\frac{a_{{\rm ini},i}}{a_{{\rm end},i}}\frac{a_{{\rm end},i}}{a_{{\rm ini},{i+1}}}\frac{H_{i}}{H_{i+1}}=c_{i+1}e^{-N_{i}}\Big(\frac{H_{i}}{H_{i+1}}\Big)^{\frac{1+3w_{i}}{3(1+w_{i})}}.\nn\ea
Similarly applying the TCC, we find an upper on $H_i$
\ba\label{Hi} \Big(\frac{H_{i}}{\mpl}\Big)^{2+\frac{2}{1+3w_{i}}}<{\textstyle\big[\frac{c_{i}}{c_{i+1}}\big]^{\frac{3(1+w_{i})}{1+3w_{i}}}}\frac{H_{i+1}}{\mpl}.\ea

In particular, in the large $m$ limit, we find the upper bound on the observable inflation as
\be H_1<\mpl{\textstyle\prod_{i=1}^{m}\big[\frac{c_i}{c_{i+1}}\big]^{\frac{3(1+w_i)}{4+6w_i}\big[\frac{1+3w_i}{4+6w_i}\big]^{i-1}}},\ee
where we define the parameter $c_{m+1}=1$. To see the importance of the above bound, we consider a special case with $c_i=10^3$ for $i=2,\dots,m$. Assuming simple post-inflationary evolutions $w_i=w$ we find
\be H_1<\mpl(10^{-3}c_1)^{1-\big[\frac{1+3w}{4+6w}\big]^{m-1}}\ee

\subsubsection*{A simple scenarion with high-scale inflation}
If we take $c_1\approx1$ and assume matter-domination in between inflations $w=0$, then
\be\label{H1} H_1<10^{-3(1-4^{1-m})}\mpl\rightarrow 10^{15}{\rm GeV} ,\ee
and \eqref{min-e-folds} implies we need at least 48.8 e-folds. 
Applying \eqref{Hm} and \eqref{Hi} repeatedly we find
\ba H_m&<&10^{\sfrac{9}{4}}H_0^{\sfrac{3}{4}}T_0^{-\sfrac{3}{4}}\Trad^{-\sfrac{1}{4}}\mpl^{\sfrac{5}{4}}\cr
H_{m-1}&<&\mpl^{\sfrac{3}{4}}H_{m}^{\sfrac{1}{4}}\cr
&\vdots&\cr
H_2&<&\mpl^{\sfrac{3}{4}}H_1^{\sfrac{1}{4}}\cr
H_1&<&\mpl^{\sfrac{3}{4}}H_{2}^{\sfrac{1}{4}}10^{-\sfrac{9}{4}}.\ea
Then, the Hubble rate of the final inflation is
\ba H_m< 
10\ {\rm TeV},\ea
(for $\Trad=10$ MeV) and the number of e-folds is $ N_m< 33.1$. 
Note that we would get the standard result of single inflation ($c_m=1$) so that $H_m<50$ GeV and $N_m<38.4$. 

The Hubble rate for the preceding one is
\be H_{m-1}<
6.1 \times10^{13} {\rm GeV},\ee
and the number of e-folds is bounded as $N_{m-1}< 10.6$. So far, we get $N<33.1+10.6=43.7$ e-folds. Note that, if this stage were the observable inflationary stage ($c_{m-1}=1$), then we would have 
\be H_{m-1}<
3.4\times10^{11} {\rm GeV},\ee
and the number of e-folds would be bounded as $N_{m-1}<15.8$. Thus the total number of e-folds is bounded by $33.1+15.8=48.9$.

For the previous inflation we find
\be H_{m-2}<
1.7 \times 10^{17} {\rm GeV},\ee
and the number of e-folds $N_{m-2}<2.6$. It violates the bound from \eqref{H1}.
Thus, it must be the observable (the first) inflation ($c_{m-2}=c_1=10^3$) with 
\be H_{m-2}<
9.6\times 10^{14} {\rm GeV},\ee
and the number of e-folds $N_{m-2}=N_1<7.8$. The total number of e-folds $N<33.1+10.6+7.8=51.5$ e-folds. This stage of inflation last for more than 7 e-folds and leave imprints on the CMB for scales spanning 3 orders of magnitude.
Moreover, if the primordial gravitational waves are discovered in future measurements, then it can be accommodated in multiple-stage inflation. 

\subsection*{III. Initial condition for the observable inflation}
The scale of the observable inflation is at least three orders of magnitude below the Planck scale. If we define $L_{\rm ini}$ as the physical size of the smooth patch at the onset of inflation which is of order $\Hi$, and $L_{\rm Pl}$ the corresponding length scale at the Planck time, then 
\be \frac{L_{\rm ini}\Hi}{L_{\rm Pl}H_{\rm Pl}}=\Big(\frac{\Hi}{H_{\rm Pl}}\Big)^{\frac{1+3\tilde w}{3(1+\tilde w)}}.\ee
In the above, $\tilde w$ is the EoS parameter of the fluid which takes over the Universe's dynamics when it has emerged from the Planck era until the dawn of observable inflation. If we take $\Hi\sim 10^{14}$ GeV and $H_{\rm Pl}\ls \mpl$ then (unless $w=-\sfrac{1}{3}$) the above equation implies that the physical volume of the Universe at the Planck time encompasses at least $10^4$ (for $w=0$ and $10^6$ for $w=\sfrac{1}{3}$) Hubble spheres with radius around Planck length. This is the initial condition for the observable inflation which must be granted if it is supposed to solve the initial condition problem of hot big bang model.

This problem can be alleviated via the very mechanism that solves the horizon problem of big bang model. After the Planck era, the universe follows an inflationary epoch with Hubble parameter close to the Planck scale. However, this inflation is not observable as we cannot observe modes that leave the horizon during this preiode. Besides, this cannot be prolonged as constrained by the TCC. 
In order to have a lower scale observable inflation with initial conditions set naturally at the Planck time without any fine-tuning we expect
\be 1\le\frac{L_{\rm ini}\Hi}{L_{\rm Pl}H_{\rm Pl}}=\frac{a_{\rm ini}\Hi}{a_{\rm Pl}H_{\rm Pl}}.\ee
Several epochs of inflation can happen in between the Planck scale one and the observable one. With a more convenient notation $H_{\rm Pl}=\tilde H_1$ and $a_{\rm Pl}=a_{{\rm ini},1}$, for $n$ stages of inflation we find
\ba 1\ge\frac{a_{{\rm ini},1}\tilde H_1}{a_{{\rm ini}}\Hi}=e^{-\sum_{i=1}^n\tilde N_i}\frac{\tilde H_1}{\Hi}\prod_{i=1}^{n-1}\frac{a_{{\rm end},i}}{a_{{\rm ini},i+1}}\frac{a_{{\rm end},n}}{a_{\rm ini}}.\ \ea
Assuming a simple inter-inflationary epoch we find
\ba\label{min-efolds} e^{\sum_{i=1}^n\tilde N_i}&\ge&\frac{\tilde H_1}{\Hi}\Big(\frac{\Hi}{\tilde H_n}\Big)^{\frac{2}{3(1+\tilde w_n)}}\prod_{i=1}^{n-1}\Big(\frac{\tilde H_{i+1}}{\tilde H_i}\Big)^{\frac{2}{3(1+\tilde w_i)}},\cr
&\rightarrow&\Big(\frac{\tilde H_1}{\Hi}\Big)^{\frac{1+3\tilde \omega}{3(1+\tilde w)}},\ea
where in the last line we assumed $w_i=w$. Therefore, 
\be \label{min-pre-inf} e^{\sum_{i=1}^n\tilde N_i} \gs 21.5\ ({\rm for}\ w=0)\quad ({\rm or}\ 100\ {\rm for}\ w=\sfrac{1}{3}),\ee
and thus $\tilde N_{\rm tot}\gs 3$ for matter-domination in intermediate stages (we would find around 4.6 e-folds for  radiation domination). 

To find the Hubble rates we parametrize the ratio of comoving horizon at the onset if each inflation ($i=1,2,\dots,n$) to the Planck horizon as
\be \frac{a_{{\rm ini},i}\tilde H_i}{a_{{\rm ini},1}\tilde H_1}=\tilde c_i\ge1.\ee
Then,
\ba\label{init-cond} \tilde c_{i+1}=\frac{a_{{\rm ini},i+1}\tilde H_i}{a_{{\rm ini},{1}}\tilde H_{1}}&=& c_{i}\frac{a_{{\rm ini},i+1}\tilde H_{i+1}}{a_{{\rm ini},{i}}\tilde H_{i}} \cr 
&=&c_{i}e^{N_{i}}\Big(\frac{\tilde H_{i+1}}{\tilde H_{i}}\Big)^\frac{1+3\tilde w_{i}}{3(1+\tilde w_{i})}.\ea
Applying the TCC, we find a lower bound on the Hubble scale of the succeeding inflation as  
\ba \tilde H_{i+1}> {\textstyle\big[\frac{\tilde c_{i+1}}{\tilde c_{i}}\big]^\frac{3(1+\tilde w_{i})}{1+3\tilde w_{i}}}\mpl\Big(\frac{\tilde H_{i}}{\mpl}\Big)^{2+\frac{2}{1+3\tilde w_{i}}}.\quad \ea
If we assume the same EoS parameter for all inter-inflationary stages and parametrize $\tilde H_1=\alpha^{-1}\mpl$ ($\alpha>1$) then we find the scale of succeeding inflation as follows
\be\label{tildeHi} \tilde H_i>\tilde C_i \alpha^{-\big[\frac{4+6\tilde w}{1+3w}\big]^{i-1}}\mpl,\ee
where
\be \tilde C_i = {\textstyle\prod}_{j=2}^{i}\Big[\Big(\frac{\tilde c_j}{\tilde c_{j-1}}\Big)^{\frac{3(1+\tilde w)}{1+3\tilde w}}\Big]^{(\frac{4+6 \tilde w}{1+3\tilde w})^{j-2}}. \ee
The number of e-folds for each stage of inflation is bounded by 
\be\tilde  N_i<\Big[\frac{4+6\tilde w}{1+3\tilde w}\Big]^{i-1}\ln\alpha-\ln c \tilde C_i,\ee
for some arbitrary constant $c>1$ which turns the inequality in \eqref{tildeHi} into equality.
The total number of e-folds, including the first one, is
\be \tilde N<\frac{\big(\frac{4+6w}{1+3w}\big)^n-1}{1+\frac{2-2w}{1+3w}}\ln\alpha-\ln c\tilde C,\ee
where ${\textstyle \ln\tilde C={\sum_{i=2}^n}{\sum_{j=2}^i}\frac{(3+3\tilde w)(4+6w)^{j-2}}{(1+3w)^{j-1}}\ln\frac{\tilde c_j}{\tilde c_{j-1}}}$.

\subsubsection*{A simple sequence of pre-inflations}
For matter domination $\tilde w=0$ and a simple model with $\tilde c_i\approx1$, 
assuming  $\alpha=\sqrt 3$ we find $\tilde H_1\approx0.5\,\mpl$, $\tilde H_2>0.1\, \mpl$ and $\tilde H_3>10^{-4}\mpl$ which touches the {
\sl Planck\,2018} bound. Consequently, the TCC implies
$\tilde N_1<0.5, \tilde N_2<2.2, \tilde N_3<8.8$. Therefore, in order to set the initial conditions of the observable inflation, there need to be at least 3 epochs of inflation which in total provide around 3 e-folds. For radiation domination $\tilde w=\sfrac{1}{3}$ we find $\tilde H_2>0.2\,\mpl$ ($\tilde \tilde N_2<1.6$), $\tilde H_3>0.007\, \mpl$ ($\tilde N_3<4.9$). Although three inflations satisfy the lower bound \eqref{min-pre-inf}, in order to satisfy the observational bound from non-detection of gravitations waves, we need at at least one more inflation before the observable one. 

\subsection*{IV. Power-law assisting inflations}
The lates results from the {\sl Planck} satellite favors plateau-like (observable) inflation. However, as the other stages of inflations have no footprint on the CMB, they could be power-law in nature. In this section, we examine this scenario for early inflations and the late ones where the Universe is dominated by a fluid with EoS parameter $\hat w<-\sfrac{1}{3}$. In this case, the Hubble rate the the beginning of inflation $H_{\rm ini}$ is not equal to the rate at the end of inflation $H_{\rm end}$. Therefore, using the dynamical equations in an expanding background, the TCC reads as follows
\be\label{modified-TCC} e^{N_i} \equiv \frac{a_{{\rm end},i}}{a_{{\rm ini},i}}<\frac{\mpl}{H_{{\rm end},i}}=\Big(\frac{\mpl}{H_{{\rm ini},i}}\Big)^{-\frac{2}{1+3\hat w_i}}.\ee
The results of this section match the results of the previous section for $\hat w=-1$.It is convenient to define the following parameters 
\ba \gamma_i&=&1-\frac{1+\hat w_i}{1+w_i},\\
p_i&=&1-\frac{2}{1+3\hat w_{i}}+\frac{2}{1+3 w_{i}},\ea
where $1\le\gamma<0$ and $p>2$ and it can be large in the limit $\hat w\rightarrow-\sfrac{1}{3}$. 

\subsubsection*{Inflation in early times}
With power-law inflation, the minimum number of e-folds to set the initial smooth path of the observable inflation is computed as 
\ba e^{\sum_{i=1}^nN_i\gamma_i}&\ge&\frac{\tilde H_{{\rm ini},1}}{\Hi}\Big(\frac{\Hi}{\tilde H_{{\rm ini},n}}\Big)^{\frac{2}{3(1+\tilde w_n)}}\cr&&\qquad\times\prod_{i=1}^{n-1}\Big(\frac{\tilde H_{{\rm ini},i+1}}{\tilde H_{{\rm ini},i}}\Big)^{\frac{2}{3(1+\tilde w_i)}},\ea
that can be compared to \eqref{min-efolds}. For $\tilde w_i=w$ we find
\be N_{\rm tot}\ge\gamma^{-1}\frac{1+3\tilde w}{3(1+\tilde w)}\ln\Big(\frac{\tilde H_{{\rm ini},1}}{\Hi}\Big).\ee
For instance, for $\hat w=-\sfrac{2}{3}$, in matter domination we obtain $N_{\rm tot}>4.6$ and in radiation domination we find $N_{\rm tot}>6.9$.
A stronger lower bound is obtained in the limit $\hat w\rightarrow -\sfrac{1}{3}$ from below and $\tilde w\rightarrow -\sfrac{1}{3}$ from above.

To find the Hubble rates of power-law inflations, we parametrize the ratio of comoving Hubble horizons as
\ba \tilde c_{i+1}&=&\frac{a_{{\rm ini},i+1}\tilde H_{{\rm ini},i+1}}{a_{{\rm ini},{1}}\tilde H_{{\rm ini},1}}= c_{i}\frac{a_{{\rm ini},i+1}\tilde H_{{\rm ini},i+1}}{a_{{\rm ini},{i}}\tilde H_{{\rm ini},i}} \cr 
&=&c_{i}e^{\gamma_iN_{i}}\Big(\frac{\tilde H_{{\rm ini},i+1}}{\tilde H_{{\rm ini},i}}\Big)^\frac{1+3\tilde w_{i}}{3(1+\tilde w_{i})}.\ea
Therefore, using the TCC relation \eqref{modified-TCC} we find
\ba \tilde H_{{\rm ini},i+1}> {\textstyle\big[\frac{\tilde c_{i+1}}{\tilde c_{i}}\big]^\frac{3(1+\tilde w_{i})}{1+3\tilde w_{i}}}\mpl\Big(\frac{\tilde H_{{\rm ini},i}}{\mpl}\Big)^{p_{i}}.\quad \ea

We note that a single epoch of power-law inflation close to the Planck scale, with EoS $\hat w$ can prepare the smooth patch of the observable inflation with Hubble rate 4 orders of magnitude less than the Planck scale. 
We parametrize $\tilde H_{{\rm ini},i}=\alpha^{-1}\mpl$ and assum $w=0$ then
\be \hat\omega=-\frac{1}{3}-\frac{2\ln\alpha}{3\big[\ln(\mpl\Hi^{-1})-3\ln\alpha\big]},\ee
and the number of e-folds is $N<\ln(\mpl\Hi^{-1})-3\ln\alpha$. For $\alpha=\sqrt3$ and $\Hi\sim10^{14}$ GeV we find $N<7.6$ (it never becomes more than $9.2$) $\hat w=-\sfrac{1}{3}-\sfrac{2\log\sqrt3}{3(4-3\log\sqrt3)}$.

\subsubsection*{Late times}
To accommodate the present horizon with plateau-like observable inflation followed by multiple power-law ones we find a lower bound on the number of e-folds
\ba 1&\ge&e^{-\sum_{i=1}^mN_i\gamma_i}\frac{T_0}{\Trad}\frac{\Hi}{H_0}\Big(\frac{\Hrad}{H_{{\rm ini},m}}\Big)^{\frac{2}{3(1+w_m)}}\cr &&\qquad\qquad\qquad\qquad\times\prod_{i=1}^{m-1}\Big(\frac{H_{{\rm ini},i+1}}{H_{{\rm ini},i}}\Big)^{\frac{2}{3(1+w_i)}}.\ \ea
For the simple case $w_i=w$ we find the total number of e-folds as $N_{\rm tot}= N_1+\gamma{\textstyle\sum_{i=2}^m}N_i$. It must obey \eqref{min-efolds}.

For the final ($m$ th) stage of inflation we find 
\ba\label{m-th} c_m&=&\frac{a_{{\rm ini},m}H_{{\rm ini},m}}{a_0H_0}=\frac{a_{{\rm ini},m}}{a_{{\rm end},m}}\frac{a_{{\rm end},m}}{a_{\rm rad}}\frac{a_{\rm rad}}{a_0}\frac{H_{{\rm ini},m}}{H_0}\\
&=& e^{-\gamma_mN_m}\frac{T_0}{H_0}\Big(\frac{H_{{\rm ini},m}}{\mpl}\Big)^{\frac{1+3w_m}{3(1+w_m)}}\Big(\frac{\Trad}{\mpl}\Big)^{\frac{1-3w_m}{3(1+w_m)}},\nn\ea
which after applying the TCC \eqref{modified-TCC} implies
\ba \Big(\frac{H_{{\rm ini},m}}{\mpl}\Big)^{p_m}<\big(c_m H_0 T_0^{-1}\big)^\frac{3(1+w_m)}{1+3w_m}\Big(\frac{\mpl}{\Trad}\Big)^{\frac{1-3w_m}{1+3w_m}}.\ea
The Hubble rate for preceding inflations $i=m\!-\!1,m\!-\!2,\dots,2,1$ are bounded by
\ba\label{m-late-mod} \Big(\frac{H_{{\rm ini},i}}{\mpl}\Big)^{p_i}&<&{\textstyle\big[\frac{c_i}{c_{i+1}}\big]^{\frac{3(1+w_i)}{1+3w_i}}}
\Big(\frac{H_{{\rm ini},i+1}}{\mpl}\Big).\ea
We recall that for the first stage $p_1=2+\frac{2}{1+3 w_{1}}$.

We can model a simple two-stage inflation to explain the horizon. The observable one with EoS parameter $\hat w=-1$ is followed by a power-law inflation with EoS parameter determined by \eqref{m-late-mod}  as
\be\hat w=-\frac{1}{3}-\frac{2}{3\big[\frac{3\ln(H_0T_0^{-1})+\ln(\mpl\Trad^{-1})}{-4\ln (\mpl\Hi^{-1})+3\ln c_2}-3\big]}.\ee
The number of e-folds is bounded as
\ba N_2<12\ln(\mpl\Hi^{-1})-&&9\ln c_2 \cr
-3\ln(H_0T_0^{-1})-&&\ln(\mpl\Trad^{-1}).\ea
For $H_1=\Hi\sim10^{14}$ GeV we find that $\hat w=-\sfrac{1}{3}-\sfrac{7}{69}$. The Hubble rate $H_2$ is around (less than) $10^{11}$ GeV. The number of e-folds is $N_2<105.8$.
Given that $N_1<9.2$, the total number of e-folds is bounded as $N_{\rm tot}<115$. 

We find that a simple three stage inflation, starting from the Planck era, can happen naturally  and explain the present horizon without any need of fine-tuning in initial conditions.

\subsection*{V. Conclusion}
In this paper we revisit the framework of multiple inflationary scenario in light of the TCC. We found that high-scale inflations in the early times after the Planck era can prepare the initial conditions for the observable inflation. The observable inflation, can be of any scale and touch the {\sl Planck\,2018} bound. If it is of high scale, then by the TCC, it cannot be prolonged and thus cannot solve the initial condition problem of the big bang model. However, we found that the succeeding inflations can explain the present horizon. All inflationary stages admit to the TCC. We also studied scenarios with a plateau-like observable inflation and supplementary power-law inflations. We found that in this scenario less number of inflations are needed provided that the EoS parameter during inflation is close to the critical value for the accelerated expansion. Eternal inflation is allowed in the framework of multi-inflation. We did not present any particular model and we describe the dynamics in terms of the effective EoS parameter of each epoch. 

\paragraph*{Acknowledgments}
This work is supported by the research deputy of SUT.
\ \\$^*$ Electronic address: mahdi@physics.sharif.ir

\end{document}